\documentclass[superscriptaddress,twocolumn,showpacs,amsmath,amssymb]{revtex4}
\usepackage[T1]{fontenc}
\usepackage{graphicx}
\usepackage{dcolumn}
\usepackage{bm}
\input epsf
\renewcommand{\vec}[1]{\mbox{\boldmath $#1$}}

\begin{document}

\preprint{}

\title{
$\beta$-decay half-lives
at finite temperatures for N=82 isotones}

\author{F. Minato}
\affiliation{
Innovative Nuclear Science Research Group, Japan Atomic Energy Agency,
Tokai 319-1195, Japan}

\author{K. Hagino}
\affiliation{
Department of Physics, Tohoku University,
Sendai 980-8578, Japan}

\date{\today}

\begin{abstract}
Using 
the finite temperature quasi-particle random phase approximation (FTQRPA)
on the basis of finite temperature Skyrme-Hartree-Fock + BCS method, 
we study $\beta^-$-decay half-lives for even-even 
neutron magic nuclei with N=82 in a finite temperature environment. 
We find that 
the $\beta^-$-decay half-life first decreases as the temperature increases 
for all the nuclei we study, although the thermal effect is found to be small at 
temperatures relevant to r-process nucleosynthesis. 
Our calculations indicate that the half-life 
begins to increase at high temperatures for open shell nuclei. 
We discuss this behavior in connection to the pairing phase transition.

\end{abstract}
\maketitle
%


%
\section{Introduction}

$\beta^-$-decay of neutron-rich nuclei is 
one of the important subjects for r-process nucleosynthesis.
In the r-process, 
nuclei rapidly capture neutrons and reach the neutron-rich region, 
until the timescale of neutron capture 
is comparable to that of the photodisintegration  
in the vicinity of the neutron shell gaps $N$=50, 82, and 126. 
The $\beta^-$-decay becomes important mainly at this point.
It increases the atomic number of neutron-rich nuclei,
and eventually enables them to go on capturing neutrons again 
toward heavier nuclei.
Therefore, the $\beta^-$-decay half-lives of neutron-rich nuclei determine 
the r-process time scale, and thus 
considerably influence the final abundance of elements. 
Likewise, the $\beta^+$-decay plays a decisive role 
in the evolution of rp-processes elements \cite{Wallace1981}.

Most $\beta^-$-decay rates of neutron-rich nuclei 
relevant to the r-process have not yet been measured experimentally. 
Therefore, r-process calculations have to rely on 
a theoretical estimate of $\beta$-decay half-lives. 
Several theoretical approaches have been developed so far. 
One of the most widely used theoretical methods is the gross theory \cite{GROSS},
which describes the $\beta$-decay rates with a sum rule approach supplemented 
by a statistical treatment for final states. 
Although it has enjoyed considerable success, 
it is not clear how well 
the shell and pairing effects for weakly bound systems are treated in the theory. 
Another approach is the shell model, 
which successfully reproduces the experimental half-lives of waiting-point nuclei 
at $N$=50, 82, and 126 \cite{Pinedo1999,Garcia2007,Langanke2000}.
However, a large-scale shell model calculation for a systematical study for heavy nuclei 
along the r-process path has been limited so far.  

The proton-neutron quasi-particle RPA (pnQRPA)
\cite{Krumlinde1984, Moller1990, Borzov1997,
Engel1999,Niksic2005,Tomislov2007,Borzov2003,Nabi2004,
Nabi1999,Homma1996,Staudt1989,Cha1983} is suitable for bridging the gap 
between the two approaches. 
Being the microscopic approach, the pnQRPA properly takes into account 
the shell and pairing effects, and 
moreover it is ideal for a systematic study. 
The strength that contributes to the $\beta^-$ decay mainly comes from 
a small low-energy tail of the Gamow-Teller (GT) distribution,
which is in general difficult to reproduce accurately with the pnQRPA.
However, the pnQRPA approach based on
the microscopic self-consistent mean-field framework 
has successfully reproduced the $\beta$-decay half-lives for neutron-rich isotopes 
by appropriately adjusting the proton-neutron pairing strength 
in the isospin $T$=0 channel 
\cite{Engel1999,Niksic2005,Tomislov2007,Borzov2003}.

The r-process takes place in an environment of high temperatures ($T\sim 10^9$ K)
and high neutron densities ($\rho \ge 10^{20}$ neutrons/cm$^3$).
In this environment, 
a part of excited states is thermally populated, and in principle one needs 
a finite temperature 
treatment for $\beta$-decay calculations for r-process. 
Notice that the thermal effects affect especially low-lying states, 
which are important for the $\beta$-decay. 
The thermal effects on the $\beta$-decay rates has been studied
with an independent particle model \cite{FFN} 
and with the finite range droplet model (FRDM) plus gross theory \cite{Famiano2008}.
The temperature dependence of electron capture rates 
has also been studied with large-scale shell model calculations\cite{Langanke2000} as 
well as with pnRPA with Skyrme interaction\cite{PCKV09}. 

In this paper, we assess the thermal effects on the $\beta$-decay of 
neutron-rich nuclei using the pnQRPA approach. 
A similar attempt has been done in Refs. 
\cite{Civitarese2000,Civitarese2001,Nabi2004,Nabi1999}, but 
they have used a schematic separable force for the particle-hole interaction. 
Some of them have neglected also the proton-neutron pairing correlation. 
We instead carry out our calculations 
based on the finite temperature Skyrme Hartree-Fock + BCS method, together with 
a contact force for the proton-neutron particle-particle 
interaction in pnQRPA.

The paper is organized as follows.
In Sec. II, we summarizes the theoretical method for finite-temperature QRPA. 
In Sec. III, we show the calculated results 
for the isotones with neutron magic number $N=82$,
which are relevant to the r-process nucleosynthesis.
In Sec. IV, we give a summary of the paper. 

\section{THEORETICAL METHODs}

\subsection{Finite temperature Hartree-Fock + BCS method}

In order to study $\beta$-decays at finite temperatures, we first construct 
the basis states using the finite temperature Hartree-Fock+BCS method 
\cite{FTHF, FTHFB}. 
The formalism of the finite temperature Hartree-Fock + BCS method 
is almost the same as that at zero-temperature \cite{Goodman1981,meantemp,VandB,V}, 
except for the particle number and pairing densities. 
At zero temperature, 
the single-particle occupation probability $n_i$ is given by the BCS occupancy $v^2_i$. 
On the other hand, at finite temperatures $\beta=1/kT$, 
$k$ being the Boltzmann constant, 
it is modified to, 
\begin{equation}
\begin{split}
n_i(T)&=f_i(T)+\tanh\left(\frac{\beta E_i}{2}\right)v_i^2,\\
f_i(T)&=\langle \alpha_i^\dagger \alpha_i \rangle=\frac{1}{1+\exp(\beta E_i)},
\end{split}
\label{occup}
\end{equation}
where $\alpha_i^\dagger$ and 
$f_i(T)$ are the creation operator and the occupation probability for 
a quasi-particle, respectively. 
$E_i=\sqrt{(\epsilon_i-\lambda)^2+(\Delta_i)^2}$ is the quasi-particle energy, where 
$\epsilon_i$ and $\lambda$ are the single-particle energy and Fermi energy, 
respectively, and the pairing gap $\Delta_i$ obeys the gap equation, 
\begin{equation}
\Delta_i=-\frac{1}{2}\sum_{j>0} V_{i\bar{i}j\bar{j}}
\frac{\Delta_j}{E_j}\tanh\left(\frac{\beta E_j}{2}\right).
\end{equation}
Here, $V$ is the pairing interaction and $\bar{i}$ is the time-reversed state of $i$.

With the densities obtained with the single-particle occupation 
probabilities $n_i$, 
the self-consistent solution is sought by minimizing the free energy,
\begin{equation}
F=E-TS(T),
\end{equation}
where $E$ is the Hartree-Fock energy and 
and $S(T)$ is the entropy defined as,
\begin{equation}
S(T)=-k\sum_i f_i(T)\ln f_i(T) + \Big(1-f_i(T)\Big)\ln \Big(1-f_i(T)\Big). 
\end{equation}

In the calculations shown below, we use the smooth cutoff scheme for 
the pairing active space, following Ref. \cite{meantemp,Bender1998}. 
That is, the quasi-particle energy and the gap equation are modified to 
$E_i=\sqrt{(\epsilon_i-\lambda)^2+(\gamma_i\Delta_i)^2}$ and 
\begin{equation}
\Delta_i=-\frac{1}{2}\sum_{j>0} V_{i\bar{i}j\bar{j}}
\frac{\Delta_j}{E_j}\tanh\left(\frac{\beta E_j}{2}\right)\gamma_i, 
\end{equation}
respectively. Here, the cutoff function is defined as \cite{meantemp,Bender1998}, 
\begin{eqnarray}
\gamma_i=\frac{1}{1+\exp{[(\epsilon_i-\lambda-\Delta E)/\mu]}}, 
\end{eqnarray}
with $\mu=\Delta E/10$.
The variable $\Delta E$ is determined so as to satisfy,
\begin{equation}
N_{act}=\sum_i \gamma_i=N_q+1.65N_q^{2/3},
\end{equation}
where $N_q$ is the number of particle for proton (q=p) or neutron (q=n).

In our calculation, 
we employ the zero-range density dependent force, 
\begin{equation}
V_{\rm{pair}}(\vec{r}-\vec{r}')
=V_q^0\left(1-\frac{\rho(\vec{r})}{\rho_0} \right)
\delta(\vec{r}-\vec{r}'),
\label{pairinginteraction}
\end{equation}
for the like-particle (proton-proton and the neutron-neutron) 
pairing interactions.
We neglect the proton-neutron pairing for the BCS calculation, although it is 
taken into account in the QRPA calculation, 
because we are interested in neutron-rich nuclei, rather than N$\simeq$Z nuclei, 
in which the proton-neutron pairing plays a minor role.
For the parameters for the pairing interaction in Eq. \eqref{pairinginteraction}, 
we fix $\rho_0$ to be 0.16 fm$^{-3}$ 
and adjust the strength parameter $V_q^0$ so as to reproduce
the empirical values for the pairing gap obtained 
from the three-point mass difference $\Delta^{(3)}(N+1)$ \cite{Satula1998}.

\subsection{Finite temperature quasi-particle random phase approximation (FT-QRPA)}

Collective motions of hot stable nuclei
have been studied with the finite temperature random phase approximation (FTRPA) 
\cite{ftVautherin, SagawaBertsch, Sommermann, Ring1984}.
It was firstly developed for studying the giant dipole resonance
of a hot compound nucleus formed in heavy-ion reactions. 
To discuss the property of hot exotic nuclei, 
the finite temperature quasi-particle RPA (FT-QRPA) was recently employed 
in Ref. \cite{E.Khan2004}.
The finite temperature proton-neutron QRPA 
has also been developed in Refs. \cite{Civitarese2000,Civitarese2001} for a 
separable interaction.

The applicability of the finite-temperature RPA has been assessed 
with the Lipkin-Meshkov-Glick method 
\cite{Rossignoli1998,Hatsuda1989,Vdovin1999,Hagino2009}.
These studies have shown that the 
finite-temperature RPA works satisfactorily well 
for the total strength. 
When the interaction is small so that the ground state is spherical, 
the FTRPA also yields a reasonable strength function.
Because the proton-neutron coupling is usually weak 
(that is, the isovector interaction is 
not large enough to 'deform' the ground state in the isospin space), 
we argue that the finite-temperature pnQRPA provides a reasonable tool
to discuss the thermal effects on the $\beta$-decay rate.

\begin{figure}
\includegraphics[width=0.90\linewidth,clip]{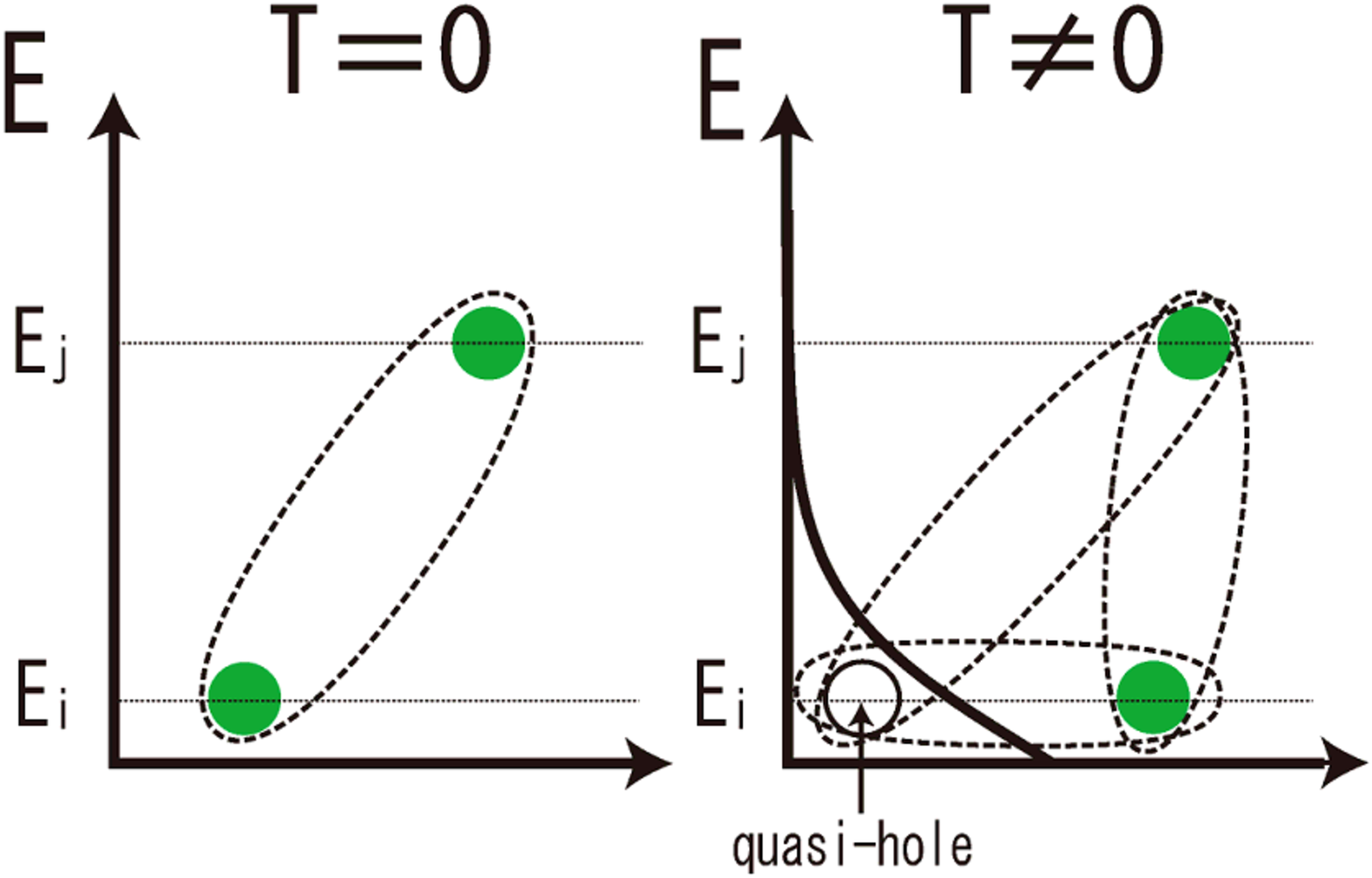}
\caption{(Color online)
A schematic illustration for the excited states in QRPA. 
At zero temperature (the left panel), excited states correspond to 
two quasi-particle (2qp) states, while 
one-quasi-particle one-quasi-hole (1qp-1qh) states 
are also involved at finite temperatures (the right panel).
The occupation probability of one quasi-particle states  
is given by $f_i(T)$ in Eq. \eqref{occup}.}
\label{transition}
\vspace{0.5cm}
\includegraphics[width=0.70\linewidth,clip]{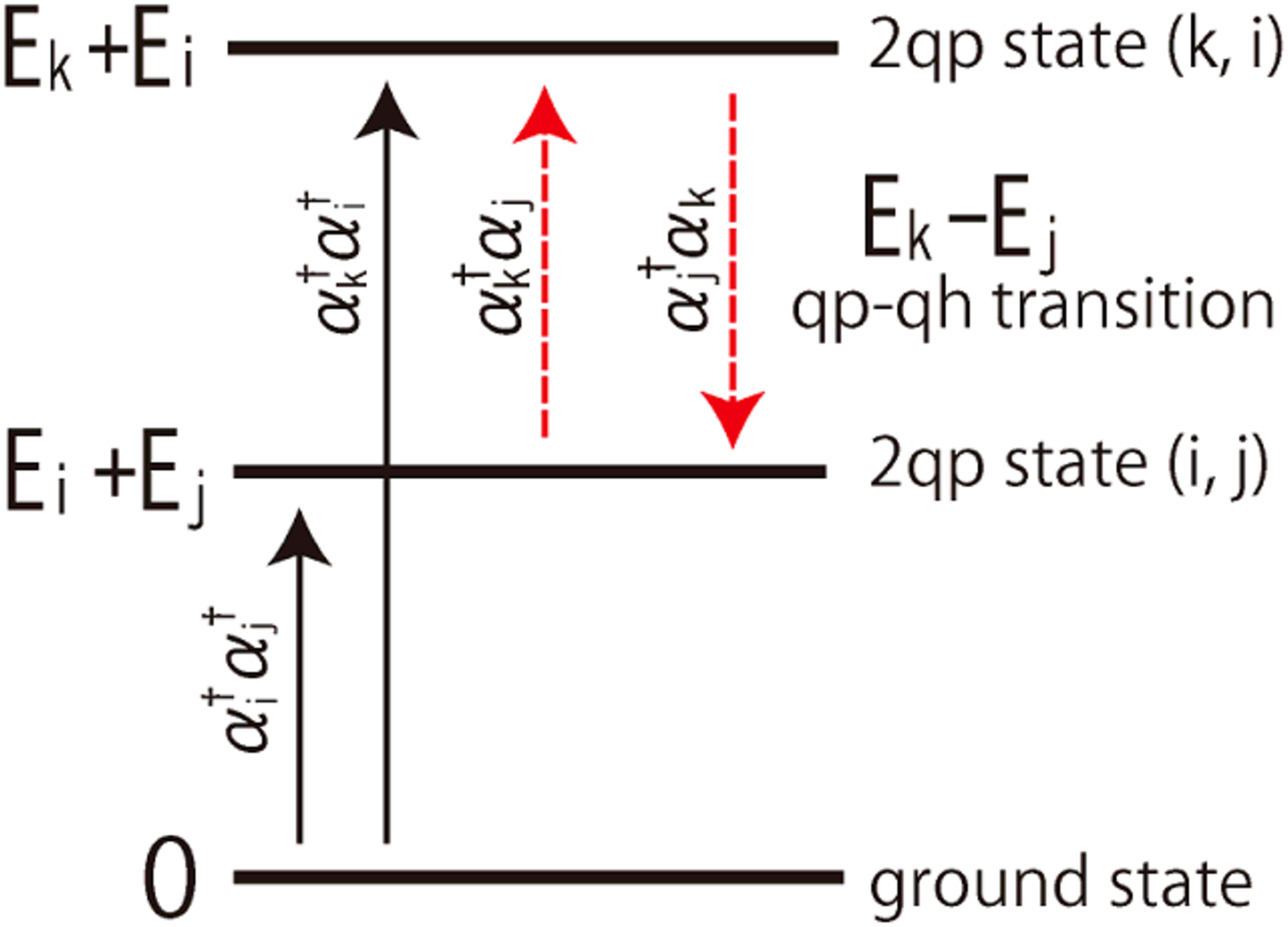}
\caption{(Color online)
A schematic illustration for excitation scheme in finite-temperature QRPA. 
Transitions from the ground state corresponds to 
a two-quasiparticle excitation, $\alpha^\dagger \alpha^\dagger$, whereas the 
transitions among two-quasiparticle states are described by the operator 
$\alpha^\dagger \alpha$. }
\label{transition2}
\end{figure}

At finite temperatures, the quasi-particle states are thermally occupied 
according to the quasi-particle occupancy $f_i(T)$ in Eq. \eqref{occup}. 
Therefore, the excitations involve both two-quasiparticle excitations 
and one-quasiparticle one-quasihole excitations, as is schematically 
shown in Figure \ref{transition}. 
This can be understood as follows (see Fig. \ref{transition2}). 
At zero-temperature, the excited states corresponds 
to two-quasi-particle states 
built on the quasi-particle vacuum. 
At finite temperatures, these excited states are thermally populated. 
The transitions among the two-quasi-particle states are then 
described by the operator $\alpha^\dagger \alpha$, {\it e.g.,} 
\begin{equation}
\alpha_k^\dagger \alpha_i^\dagger |0\rangle = 
\alpha_k^\dagger \alpha_j [\alpha_j^\dagger \alpha_i^\dagger |0\rangle]. 
\end{equation}
The energy change for this transition is 
\begin{equation}
\Delta E = (E_k+E_i)-(E_i+E_j)=E_k-E_j.
\end{equation}

The transition operator at a finite temperature thus reads \cite{Sommermann},
\begin{equation}
Q^\dagger
=\sum_{\alpha,\beta}P_{\alpha\beta}\alpha_\alpha^\dagger\alpha_\beta
                   +X_{\alpha\beta}\alpha_\alpha^\dagger\alpha_\beta^\dagger
                   -Q_{\alpha\beta}\alpha_\alpha        \alpha_\beta^\dagger
                   -Y_{\alpha\beta}\alpha_\alpha        \alpha_\beta.
\label{Transition}
\end{equation}
where $\alpha$ and $\beta$ run over proton and neutron levels, respectively. 
The first and third terms in Eq. (\ref{Transition}) correspond to 
the transitions among the two-quasiparticle states,
which vanish at zero-temperature.
The QRPA equation can be derived from the equation of motion, 
$\langle|[\delta Q,[H,Q^\dagger]]|\rangle
=E_{\rm{QRPA}}\langle|[\delta Q,Q^\dagger]|\rangle$,
where $E_{\rm{QRPA}}$ is the QRPA excitation energy and $\delta Q$ is any 
one-body operator. 
This yields, 
\begin{equation}
\left(
\begin{tabular}{rrrr}
$ \tilde{C}  $ & $ \tilde{a}$ & $ \tilde{D}  $ & $ \tilde{b}$ \\
$ \tilde{a}^T$ & $ \tilde{A}$ & $ \tilde{b}^T$ & $ \tilde{B}$ \\
$-\tilde{D}  $ & $-\tilde{b}$ & $-\tilde{C}  $ & $-\tilde{a}$ \\
$-\tilde{b}^T$ & $-\tilde{B}$ & $-\tilde{a}^T$ & $-\tilde{A}$
\end{tabular}
\right)
\left(
\begin{tabular}{c}
$\tilde{P}$\\
$\tilde{X}$\\
$\tilde{Q}$\\
$\tilde{Y}$\\
\end{tabular}
\right)
=
E_{\rm{QRPA}}\left(
\begin{tabular}{c}
$\tilde{P}$\\
$\tilde{X}$\\
$\tilde{Q}$\\
$\tilde{Y}$\\
\end{tabular}
\right)
\label{FTQRPA}
\end{equation}
where the elements of the matrices 
$\tilde{A}, \tilde{B}, \tilde{C}, \tilde{D}, \tilde{a},$ and $\tilde{b}$ 
are given by \cite{Sommermann},
\begin{equation}
\begin{split}
\tilde{A}_{\alpha\beta\alpha'\beta'}
&=\sqrt{1-f_\alpha-f_\beta}A_{\alpha \beta \alpha' \beta'}'
\sqrt{1-f_{\alpha'}-f_{\beta'}}\\
&+(E_\alpha+E_\beta)\delta_{\alpha\alpha'}\delta_{\beta\beta'}\\
\tilde{B}_{\alpha \beta \alpha' \beta'}&=
\sqrt{1-f_\alpha-f_\beta}B_{\alpha \beta \alpha' \beta'}
\sqrt{1-f_{\alpha'}-f_{\beta'}}\\
\tilde{C}_{\alpha \beta \alpha'\beta'}&
=\sqrt{f_\beta-f_\alpha}C_{\alpha\beta\alpha' \beta'}
\sqrt{f_{\beta'}-f_{\alpha'}}\\
&+(E_\alpha-E_\beta)\delta_{\alpha \alpha'}\delta_{\beta\beta'}\\
\tilde{D}_{\alpha \beta \alpha' \beta'}&
=\sqrt{f_\beta-f_\alpha}D_{\alpha \beta \alpha' \beta'}
\sqrt{f_{\beta'}-f_{\alpha'}}\\
\tilde{a}_{\alpha \beta \alpha' \beta'}&
=\sqrt{f_\beta - f_\alpha}a_{\alpha \beta \alpha' \beta'}
\sqrt{1-f_{\alpha'}-F_{\beta'}}\\
\tilde{b}_{\alpha \beta \alpha' \beta'}&
=\sqrt{f_\beta- f_\alpha}b_{\alpha \beta \alpha' \beta'}
\sqrt{1-f_{\alpha'}-f_{\beta'}},
\end{split}
\label{QRPA}
\end{equation}
with 
\begin{equation}
\begin{split}
A_{\alpha\beta\alpha'\beta'}'=
& \quad V_{\alpha\beta\alpha'\beta'}
(u_\alpha u_\beta u_{\alpha'}u_{\beta'}+v_\alpha v_\beta v_{\alpha'}v_{\beta'})\\
&+V_{\alpha \bar{\beta}'\bar{\beta}\alpha'}
(u_\alpha v_\beta u_{\alpha'}v_{\beta'}+v_\alpha u_\beta v_{\alpha'}u_{\beta'})\\
B_{\alpha\beta\alpha'\beta'}=
& \quad V_{\alpha \beta' \bar{\alpha}' \bar{\beta}}
(u_\alpha v_\beta v_{\alpha'}u_{\beta'}+v_\alpha u_\beta u_{\alpha'}v_{\beta'})\\
&-V_{\alpha \beta \bar{\alpha}'\bar{\beta}'}
(u_\alpha u_\beta v_{\alpha'}v_{\beta'}+v_\alpha v_\beta u_{\alpha'}u_{\beta'})\\
C_{\alpha\beta\alpha'\beta'}'=
& \quad V_{\alpha \beta' \beta \alpha'}
(u_\alpha u_\beta u_{\alpha'}u_{\beta'}+v_\alpha v_\beta v_{\alpha'}v_{\beta'})\\
&-V_{\alpha \bar{\beta} \bar{\beta}'\alpha}
(u_\alpha v_\beta u_{\alpha'}v_{\beta'}+v_\alpha u_\beta v_{\alpha'}u_{\beta'})\\
D_{\alpha \beta \alpha' \beta'}=
&-V_{\alpha \beta \bar{\alpha}' \bar{\beta}'}
(u_\alpha v_\beta v_{\alpha'}u_{\beta'}+v_\alpha u_\beta u_{\alpha'}v_{\beta'})\\
&+V_{\alpha\bar{\beta}'\bar{\alpha}'\beta}
(u_\alpha u_\beta v_{\alpha'}v_{\beta'}+v_\alpha v_\beta u_{\alpha'}u_{\beta'})\\
a_{\alpha\beta\alpha'\beta'}=
& \quad V_{\alpha \bar{\beta}\alpha'\beta'}
(v_\alpha u_\beta v_{\alpha'}v_{\beta'}-u_\alpha v_\beta u_{\alpha'}u_{\beta'})\\
&-V_{\alpha \bar{\beta}'\beta \alpha'}
(v_\alpha v_\beta v_{\alpha'}u_{\beta'}-u_\alpha u_\beta u_{\alpha'}v_{\beta'})\\
b_{\alpha\beta\alpha'\beta'}=
& \quad V_{\bar{\alpha}\beta \alpha'\beta'}
(v_\alpha u_\beta u_{\alpha'}u_{\beta'}-u_\alpha v_\beta v_{\alpha'}v_{\beta'})\\
&-V_{\alpha \beta'\beta\bar{\alpha}'}
(v_\alpha v_\beta u_{\alpha'}v_{\beta'}-u_\alpha u_\beta v_{\alpha'}u_{\beta'}).\\
\end{split}
\label{finiteab}
\end{equation}
Using the solution of the QRPA equation, 
the strength function $S^\pm(E)$ for the GT transition is calculated as ,
\begin{equation}
\begin{split}
&S^\pm(E_\nu)=\frac{1}{1-\exp(-\beta E_\nu)}\\
&\times
\Bigl| \sum_{\alpha > \beta}\langle \alpha|\mathcal{O}_{GT}^\pm|\beta \rangle
\Big( (u_\alpha u_\beta P_{\alpha\beta}^\nu + v_\alpha v_\beta Q_{\alpha\beta}^\nu)
\sqrt{f_\beta-f_\alpha}\Bigr.\\
& \Bigl. \qquad\qquad
     +(u_\alpha v_\beta X_{\alpha\beta}^\nu + v_\alpha u_\beta Y_{\alpha\beta}^\nu)
\sqrt{1-f_\beta-f_\alpha}
   \Big)\Bigr|^2\\
&\times
\delta(E_\alpha-E_\beta-E_\nu),
\end{split}
\label{strength}
\end{equation}
where $\mathcal{O}_{GT}^\pm=\sigma\tau^\pm$.
For $T=0$,
one can see that the Eqs. \eqref{FTQRPA} and \eqref{strength} 
are correctly reduced to the usual QRPA equation at zero temperature.

In our calculations, we use the $t_0$ and $t_3$ terms 
in the Skyrme force \cite{LM} as the residual interaction,
\begin{equation}
v(\vec{r},\vec{r}')=
-\left(
\frac{t_0}{4}+\frac{t_3}{24}\rho^\alpha(\vec{r})\right)
(\sigma\cdot\sigma) (\tau\cdot\tau) \,
\delta(\vec{r}-\vec{r}'),
\label{residual}
\end{equation}
for the Gamow-Teller transition.
For the particle-particle matrix elements (the proton-neutron isospin $T$=0 pairing) 
in Eq. \eqref{finiteab}, 
we use a $\delta$-type interaction,
\begin{equation}
V_{pn}(\vec{r},\vec{r}')=V_{pn}^0 \delta(\vec{r}-\vec{r}'). 
\label{pnresid}
\end{equation}
We can regard $V_{pn}^0$ as a free parameter 
as has been discussed in Ref. \cite{Engel1999}, 
because we do not take into account the $T$=0 pairing 
in the Hartree-Fock calculation.
The Gamow-Teller low-lying strengths are sensitive to the $T$=0 pairing,
and we adjust the value of $V_{pn}^0$ to reproduce the known experimental 
$\beta$ half-life at zero temperature 
\cite{Engel1999,Cha1983}.

We solve the QRPA equation by diagonalizing the QRPA matrix in 
Eq.\eqref{FTQRPA}. 
In order to include continuum states, we discretize them with 
a box boundary condition with the box size of 15 fm. 
We include the single-particle states 
up to $\epsilon=20$ MeV, and truncate the QRPA model 
space at the two quasi-particle energy of $E_{2qp}=70$ MeV.
Our pnQRPA calculation is not fully self-consistent, since 
we do not include all the residual interaction terms in the Skyrme functional. 
We thus scale the residual interaction 
Eq.\eqref{residual} so as to reproduce the 
spurious translational mode (that is, the isoscalar dipole mode) 
at zero energy at every temperature we consider.

\section{RESULTs}

\subsection{Temperature dependence of GT strengths for N=82 Nuclei}
\label{TDGT}

Let us now numerically solve the pnQRPA equations and discuss 
the temperature dependence of the GT strengths for even-even 
$N$=82 nuclei, $^{120}$Sr, $^{122}$Zr, $^{124}$Mo, 
$^{126}$Ru, $^{128}$Pd, and $^{130}$Cd. 
For this purpose, we mainly use the SLy5 force \cite{SLy5} for 
the Skyrme parameter set. 
We set the proton pairing strength $V_p=-1300$ MeV$\cdot$fm$^{-3}$ 
so as to reproduce the empirical pairing gap
of $^{130}$Cd, that is, $\Delta^{(3)}_p(Z+1)=0.92$ MeV.
The proton-neutron pairing strength 
in Eq. \eqref{pnresid} is adjusted to $V_{pn}^0=-360$ MeV$\cdot$fm$^{-3}$ 
so as to reproduce the experimental $\beta$-decay half-life 
of $^{130}$Cd ($0.195$ sec.) \cite{130Cd}, 
and use the same value for all the other nuclei. 

\begin{figure}
\begin{center}
\includegraphics[width=.4\textwidth,clip]{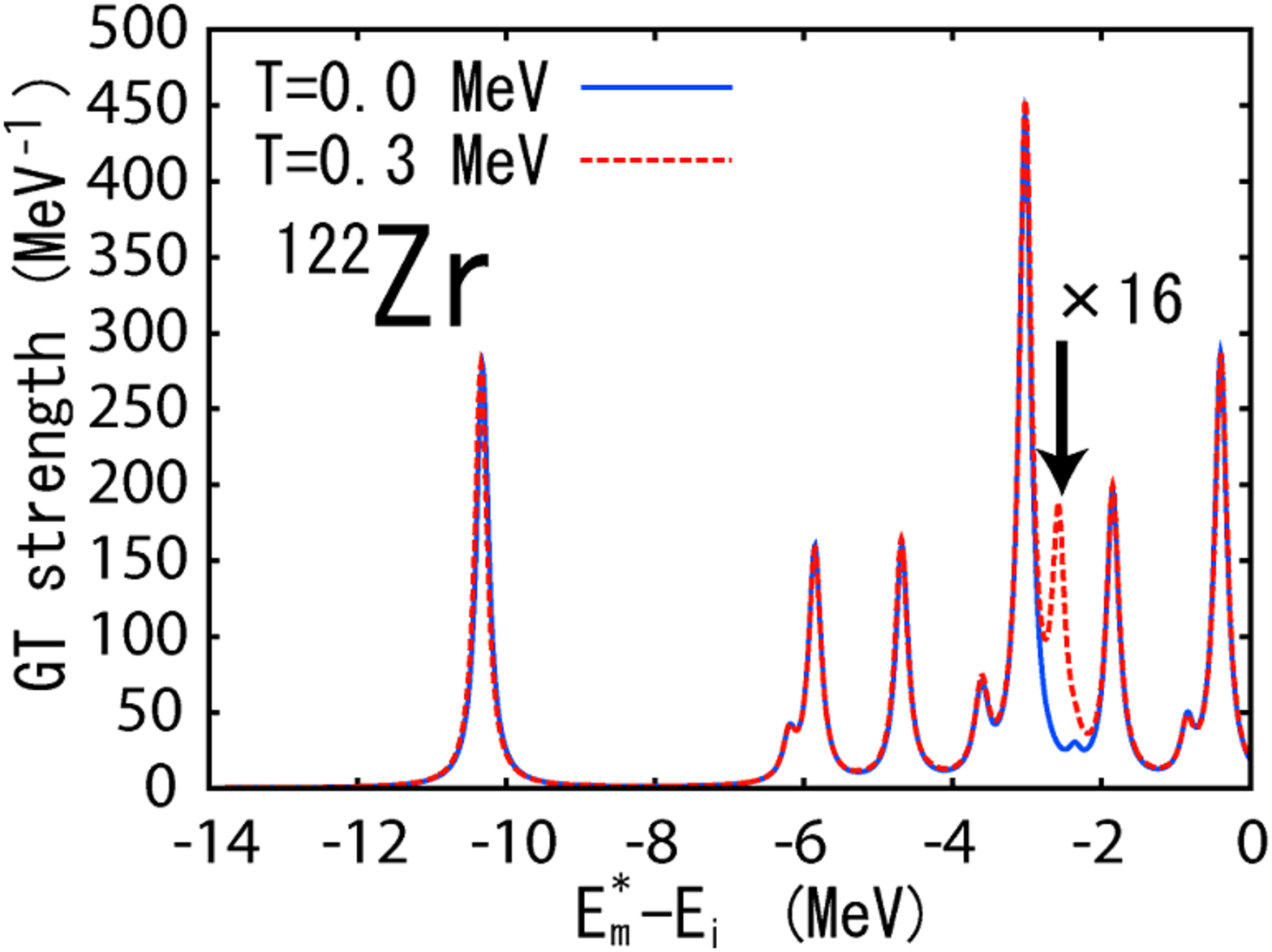}
\includegraphics[width=.4\textwidth,clip]{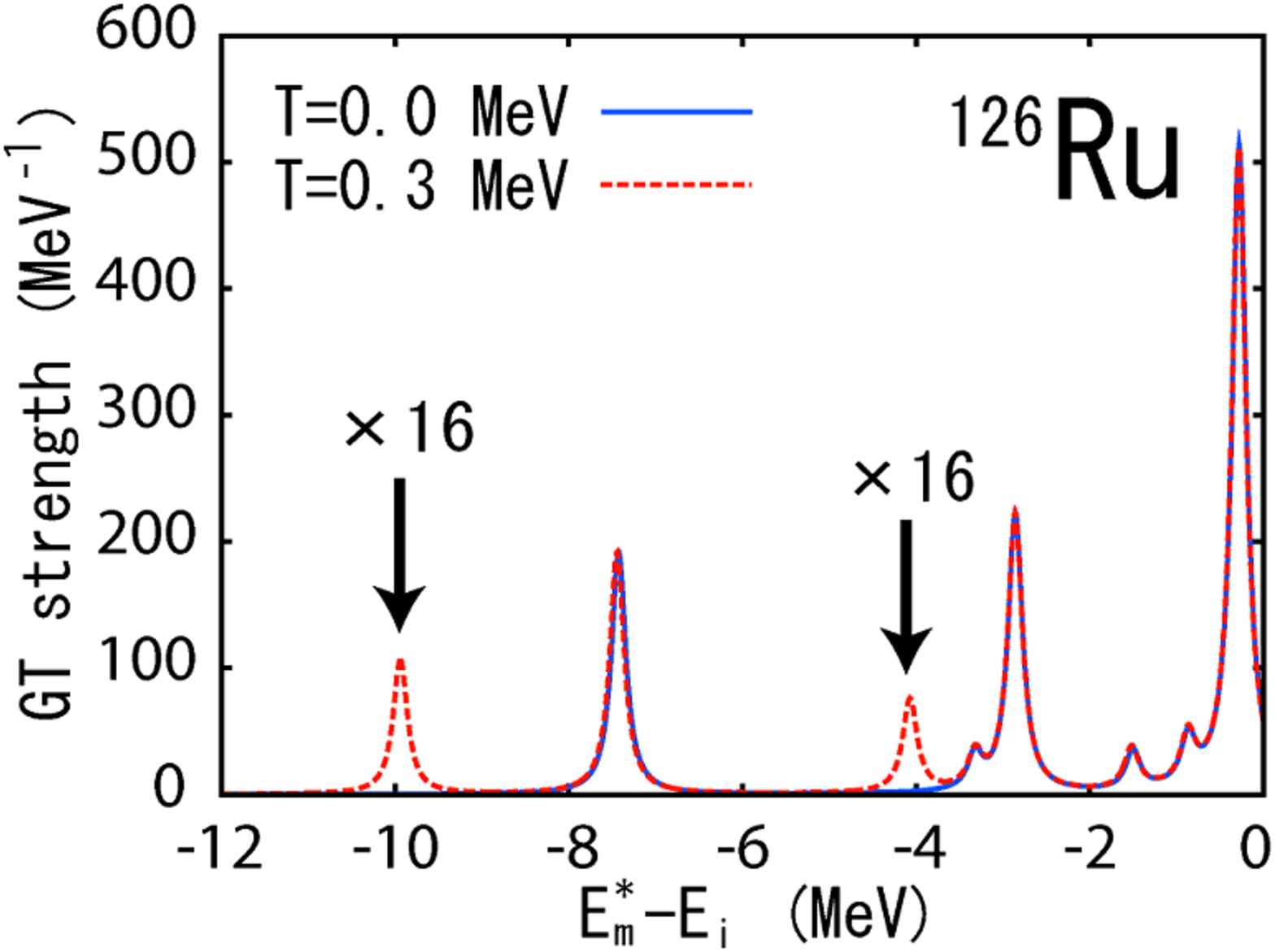}
\includegraphics[width=.4\textwidth,clip]{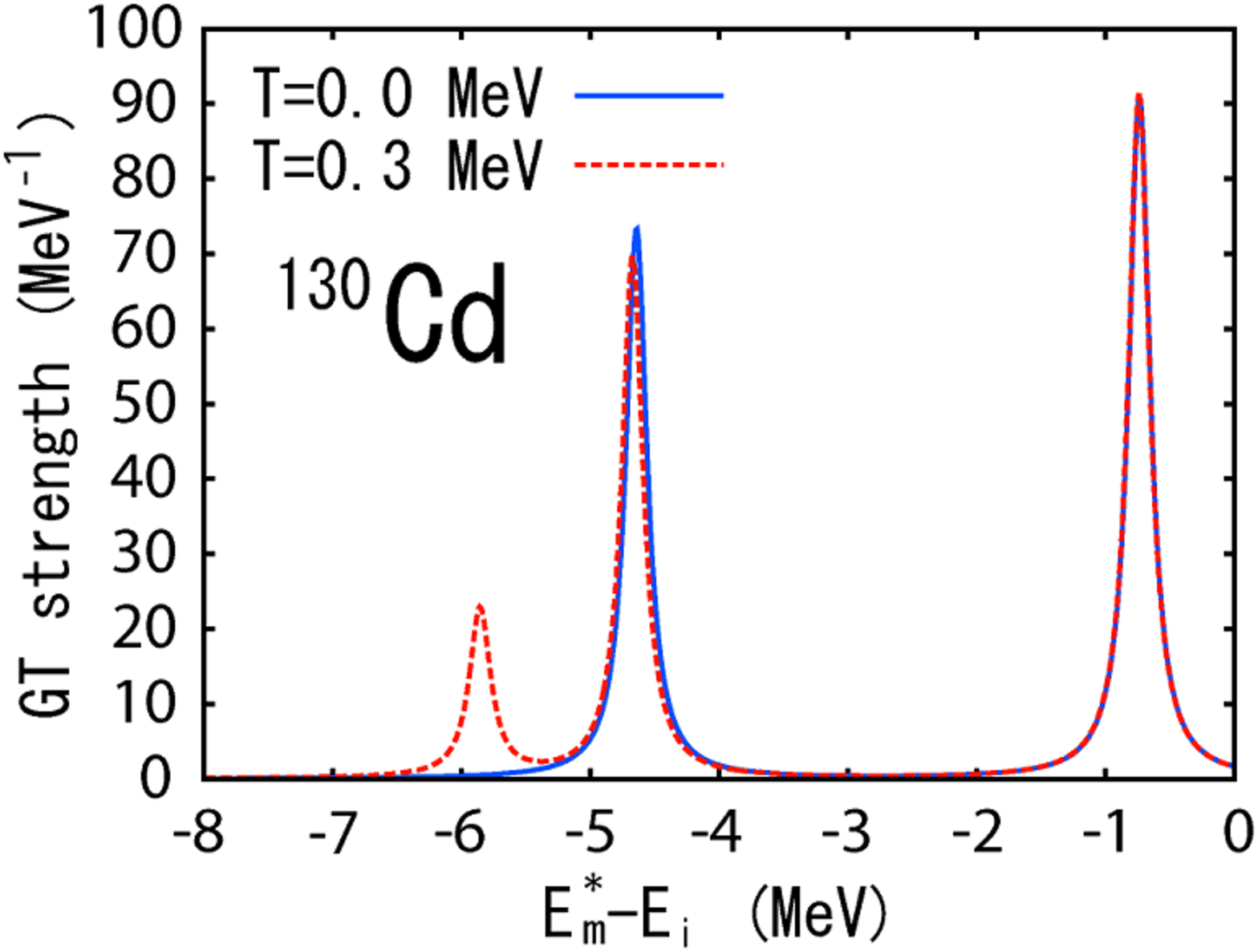}
\end{center}
\caption{(Color online)
The Gamow-Teller strength functions for the 
$^{122}$Zr, $^{126}$Ru, and $^{130}$Cd nuclei 
at $T=0.0$ (the solid line) and $T=0.3$ MeV (the dashed line).
These are plotted as a function of $E_m^* - E_i$, 
where $E_i$ and $E_m^*$ are the energy of the initial 
and final states, respectively.  
The peaks which vanish at $T=0.0$ MeV are indicated by the arrows.
For $^{126}$Ru and $^{122}$Zr, the strengths at $T=0.3$ MeV are 
scaled by a factor of 16. } 
\label{GTstrengthbeta}
\end{figure}

\begin{figure}
\begin{center}
\includegraphics[width=.40\textwidth,clip]{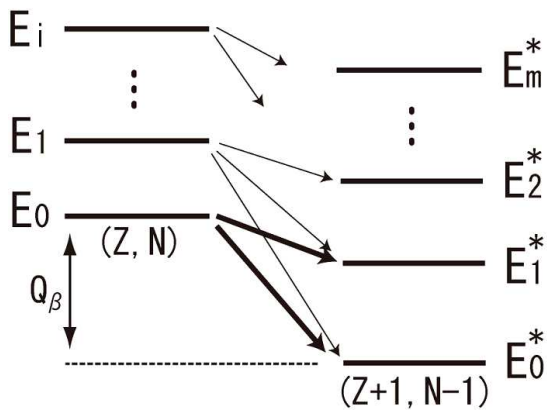}
\end{center}
\caption{A schematic illustration for 
the $\beta^-$-decay scheme at finite temperatures. 
The transitions at zero temperature 
are indicated by the thick solid arrows, while the additional 
transitions at finite temperatures by the thin solid arrows.}
\label{WeakBranch}
\end{figure}

We find that the strength function is almost the same as 
that at $T=0$ for temperatures less than $T=0.2$ MeV, 
which is considered to be the standard r-process temperature 
at the initial condition \cite{Woosley1994}.
Figure \ref{GTstrengthbeta} shows the GT strengths
at $T=0.0$ (the solid line) and $T=0.3$ MeV (the dashed line) 
for the $^{122}$Zr, $^{126}$Ru, and $^{130}$Cd nuclei 
as a function of $E_m^* - E_i$, 
where $E_i$ and $E_m^*$ are the energy of the initial 
and final states, respectively 
(See Fig.\ref{WeakBranch} and Eq. \eqref{5.6.}).
Those strength functions are smeared with the Lorentzian function with 
the width of 0.1 MeV. 
The strengths at $T=0.3$ MeV for $^{126}$Ru and $^{122}$Zr
are multiplied by a factor of 16 for the presentation purpose. 
One sees that some new peaks, indicated by the arrows, 
appear at $T=0.3$ MeV, which originate 
from the transition from the excited states.
Their strengths are of the order of 0.1 on average, 
which are approximately of 0.1 \% of the sum rule.
Despite its small value, these contributions  
to the $\beta$-decay half-life cannot be neglected 
as we will discuss in the next section.

\subsection{$\beta$-decay Half-Lives}
\label{BDHL}

We next calculate the $\beta^-$-decay half-lives. 
Since the contribution of the GT transition 
to the total $\beta^-$-decay rate 
is much larger than the Fermi transition \cite{Langanke2003}, 
we take into account only the former. 
The $\beta$-decay half-life $T_{1/2}$ can be calculated with 
the Fermi Golden rule as \cite{Engel1999, Bowler}, 
\begin{equation}
\begin{split}
\frac{1}{T_{1/2}}
&=\frac{\lambda_\beta}{\ln 2}\\
&=\frac{G_F^2}{\ln 2}\frac{g_A^2}{\hbar}
\int_0^{\infty} dE_e \sum_m S^-(E_m)
\,\rho( E_i-E_m^*,E_e ),
\end{split}
\label{betadecayrate}
\end{equation}
where $\lambda_\beta$ is the $\beta$-decay rate.
$G_F=1.1658 \times 10^{-11} \, \mathrm{MeV^{-2}}$ is the Fermi constant, and 
$g_A=G_A/G_V$ is the ratio of the 
vector and pseudo vector constants, which we set 1.26. 
The function $\rho(E,E_e)$ is the phase space factor for the outgoing
electron and anti-neutrino given by,
\begin{equation}
\rho(E,E_e)=\frac{E_e\sqrt{E_e^2-m_e^2}}{2\pi^3}(E-E_e)^2 F(Z,E_e), 
\label{PhaseSpace}
\end{equation}
where $E_e$ is the energy of the electron and $Z$ is the atomic number 
of the parent nucleus. 
$F(Z,E_e)$ is the Coulomb correction factor given by \cite{Behrens1982},
\begin{equation}
F(Z,E_e)=2(1+\gamma) (2k_eR_n)^{2(\gamma-1)} 
\Big|\frac{\Gamma(\gamma+i\nu)}{\Gamma(2\gamma+1)}\Big|^2 e^{\pi \nu},
\end{equation}
where $\gamma = (1-Z^2 \alpha^2)^{1/2} $ and $\nu=(Z \alpha E_e/p_e c)$,
$\alpha$ being the fine structure constant.
$k_e=p_e/\hbar$ is the electron wave number 
and $\Gamma(x)$ is the gamma function.
The energy $E_i-E_m^*$ in Eq. \eqref{PhaseSpace} is related to 
the pnQRPA energy $E_{\rm{QRPA}}$ as \cite{Engel1999},
\begin{equation}
\begin{split}
E_i-E_m^*
&\simeq \Delta M_{n-H}-(E_{\rm{RPA}}-\lambda_n+\lambda_p), 
\end{split}
\label{5.6.}
\end{equation}
where $\Delta M_{n-H}=0.78227$ MeV is the mass difference between a 
neutron and a hydrogen atom. 

\begin{figure*}[t]
\begin{center}
\begin{tabular}{cc}
\includegraphics[width=.42\textwidth,clip]{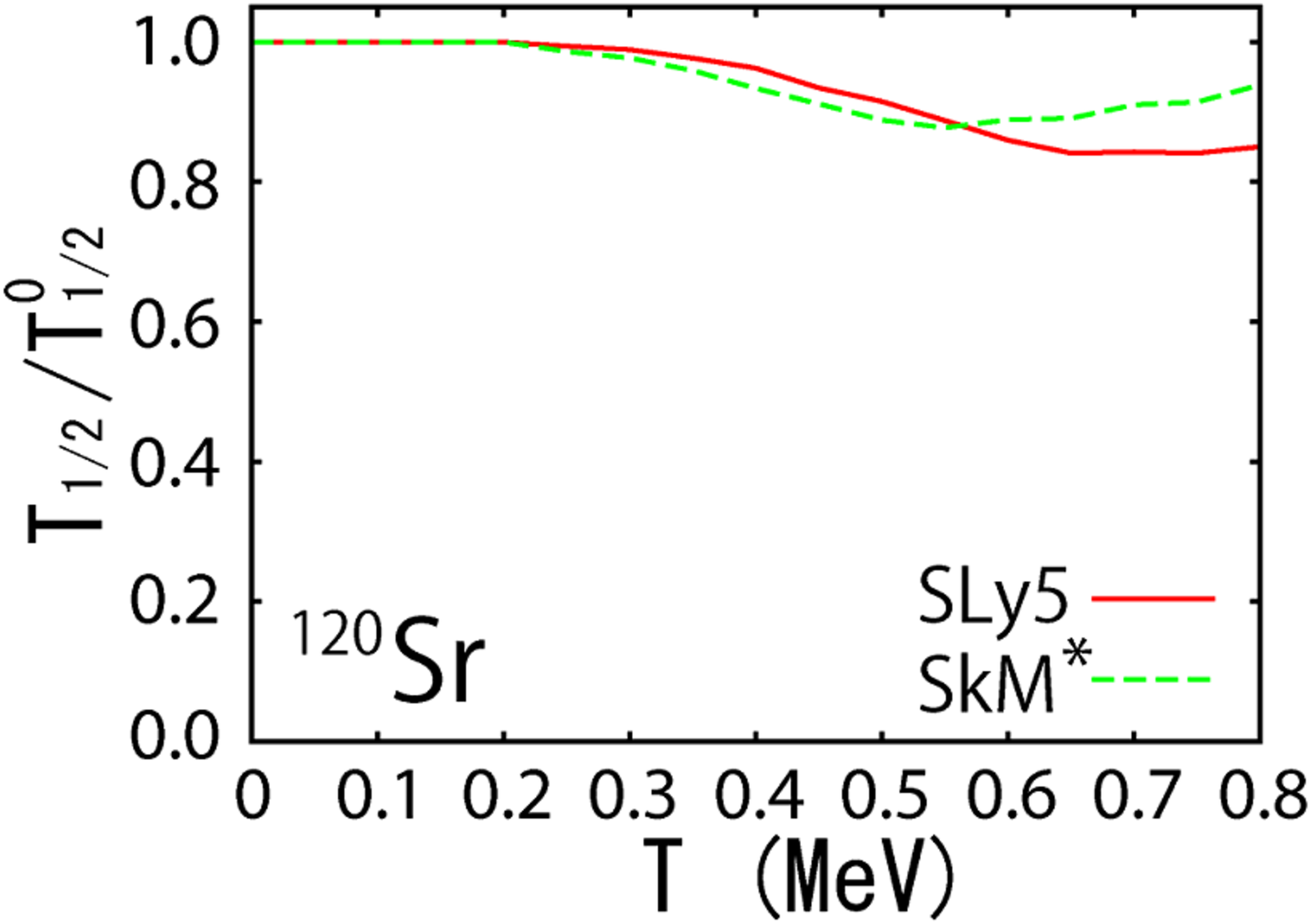}&
\includegraphics[width=.42\textwidth,clip]{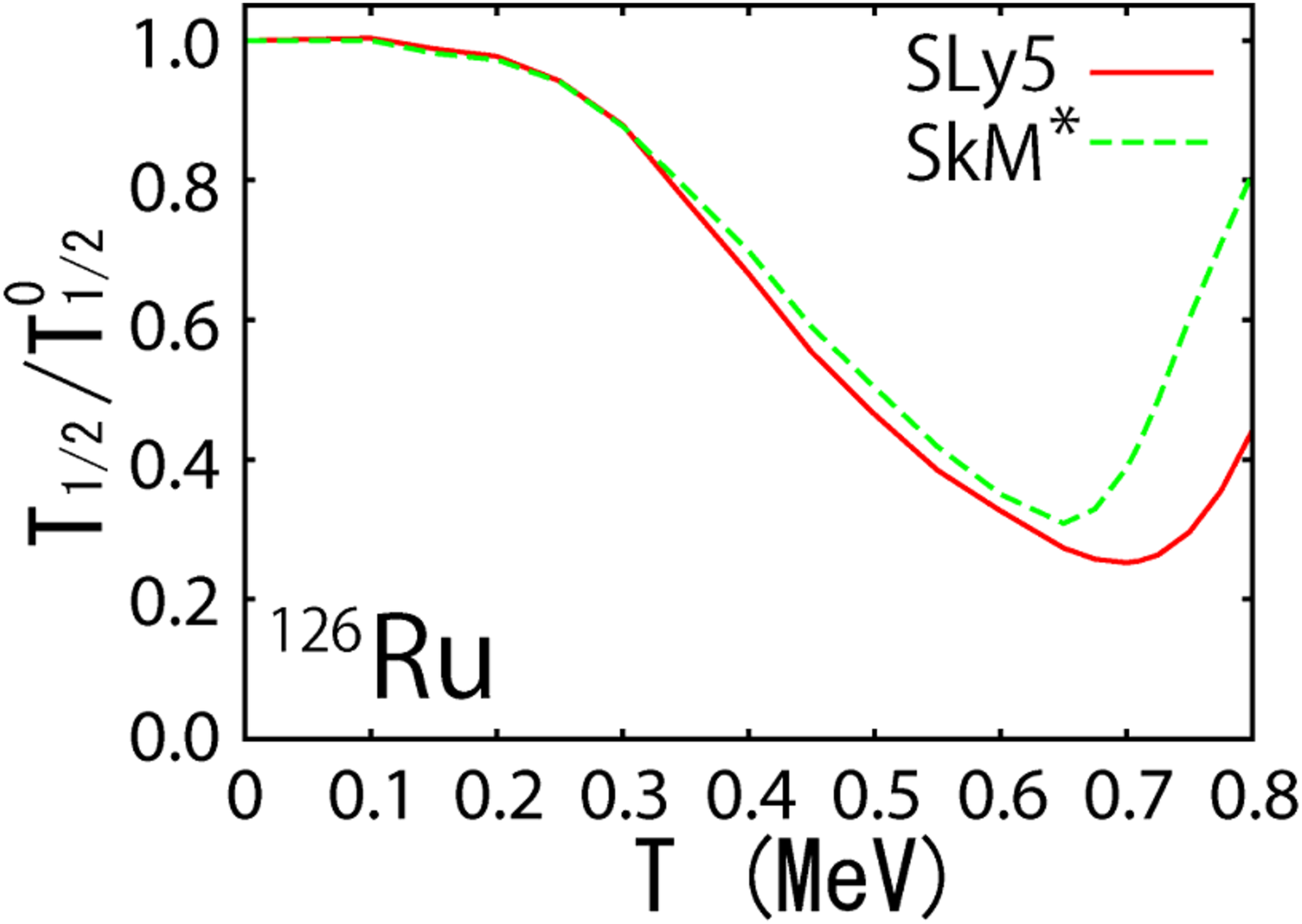}\\
\includegraphics[width=.42\textwidth,clip]{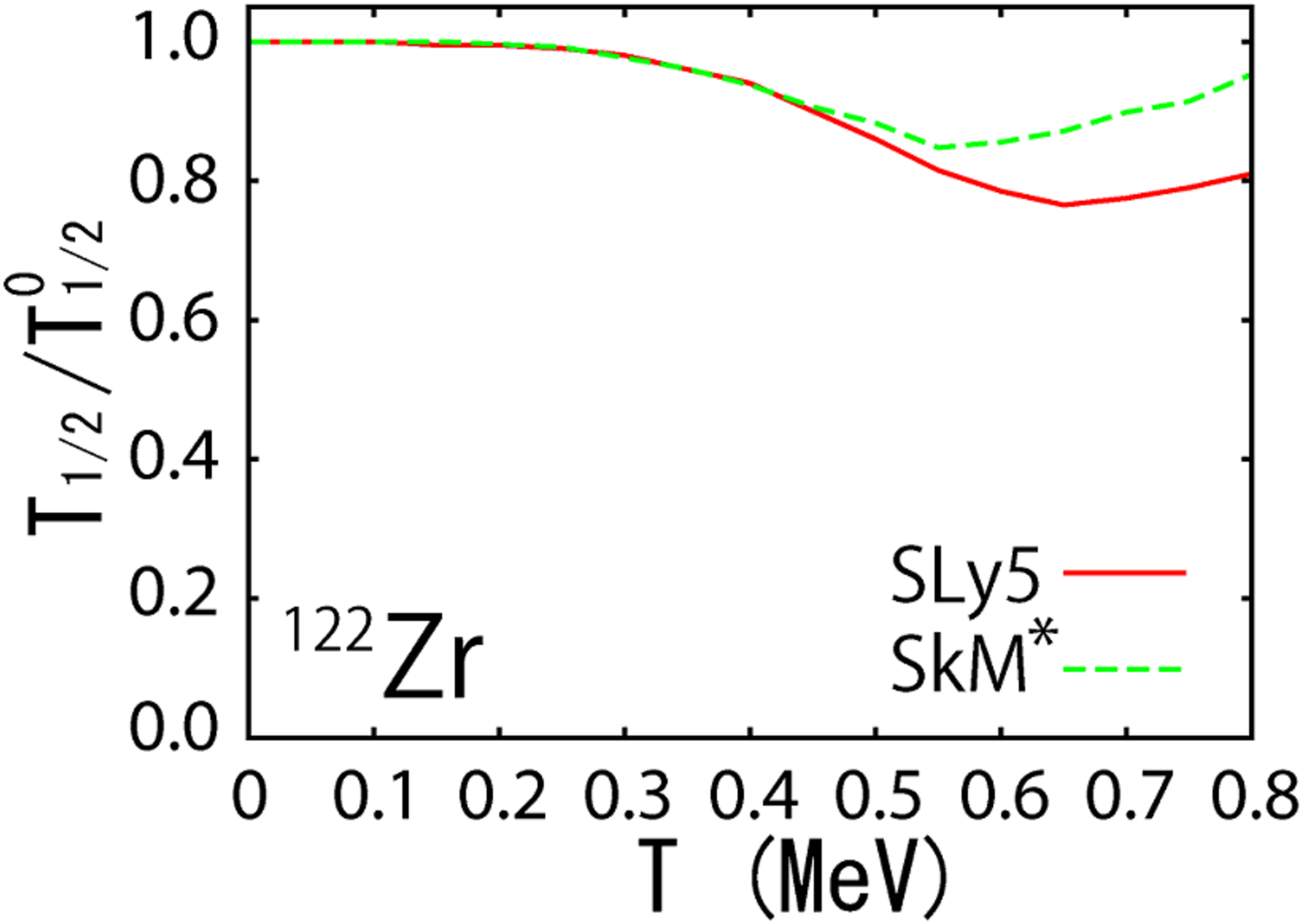}&
\includegraphics[width=.42\textwidth,clip]{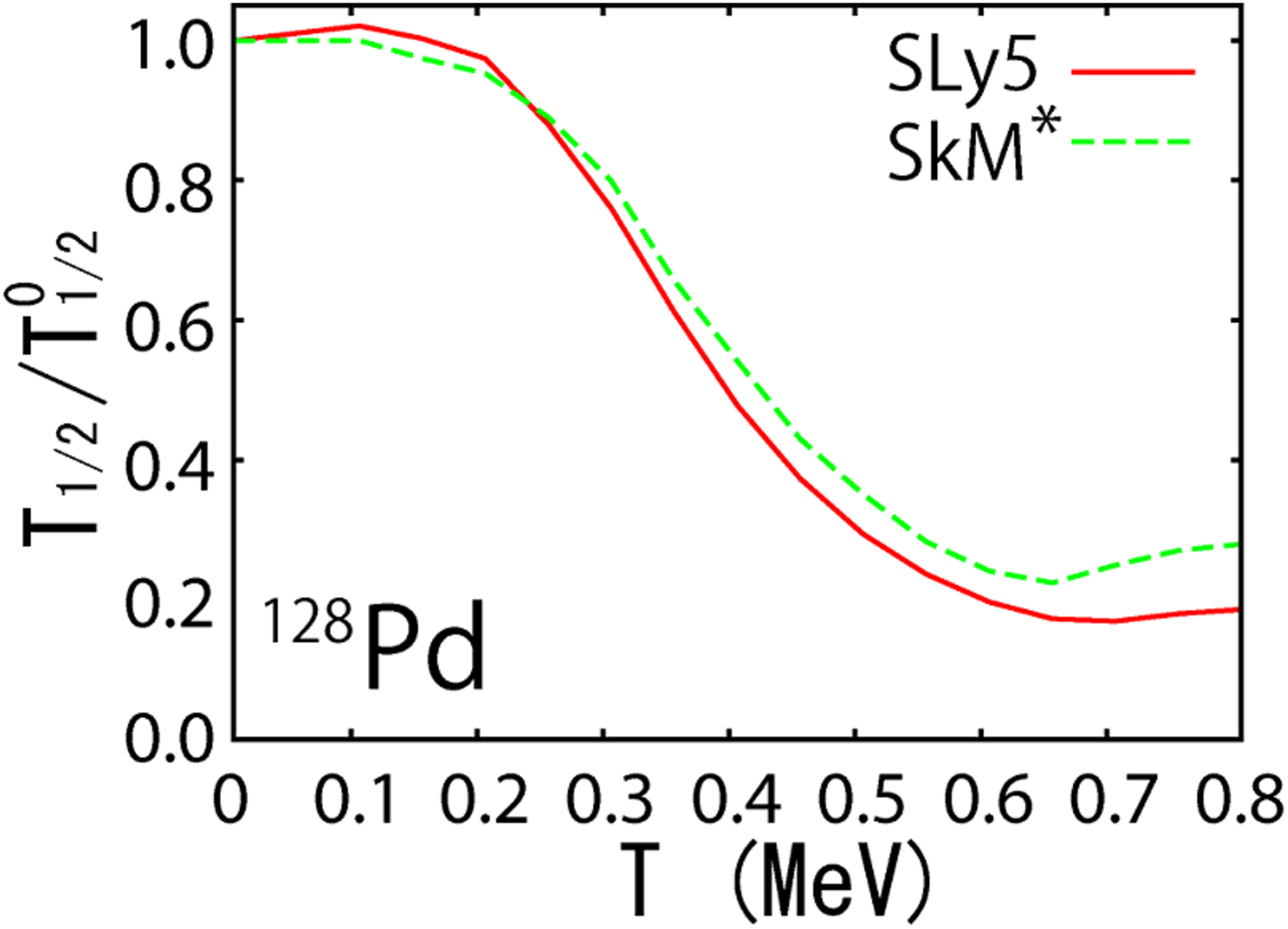}\\
\includegraphics[width=.42\textwidth,clip]{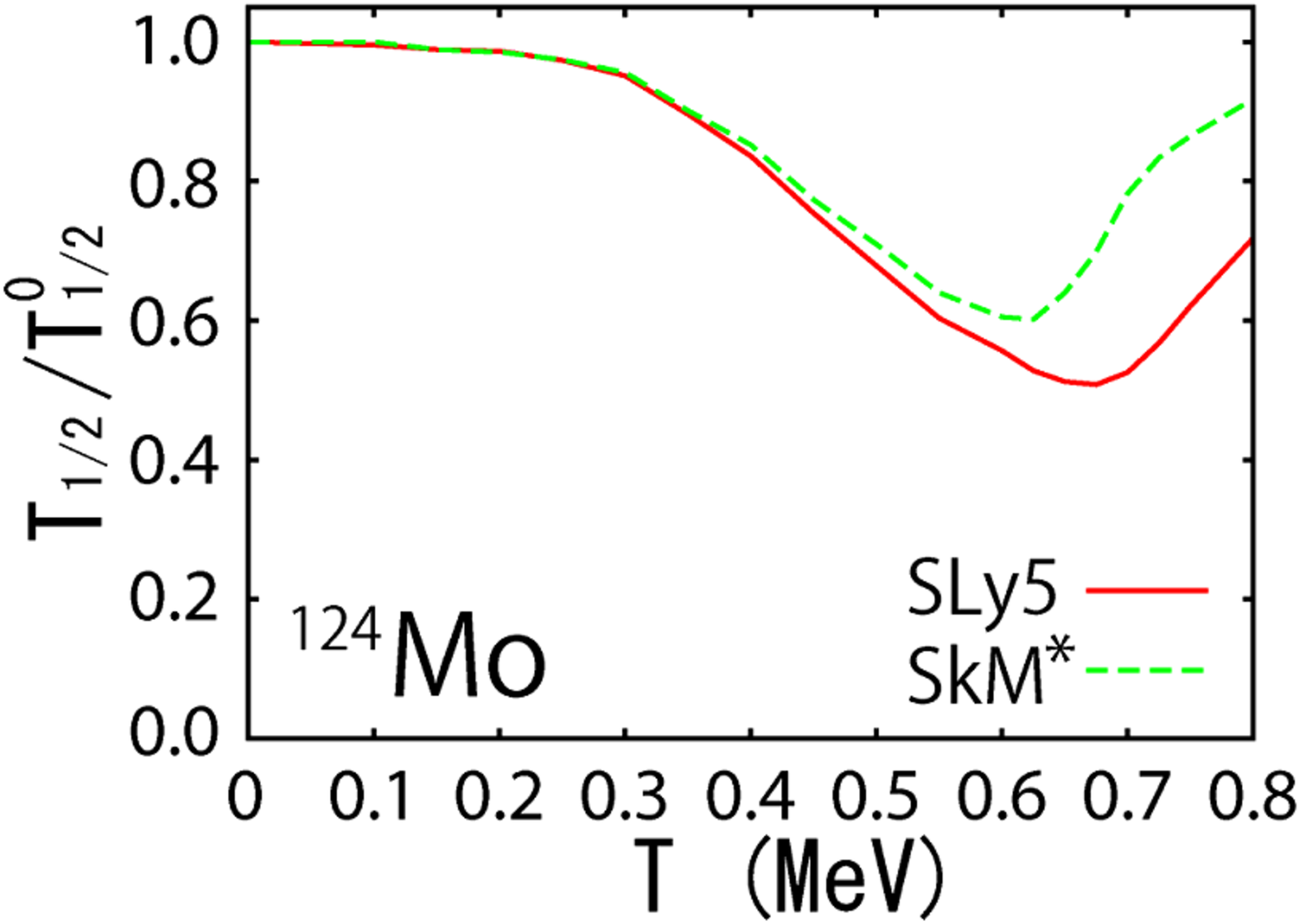}&
\includegraphics[width=.42\textwidth,clip]{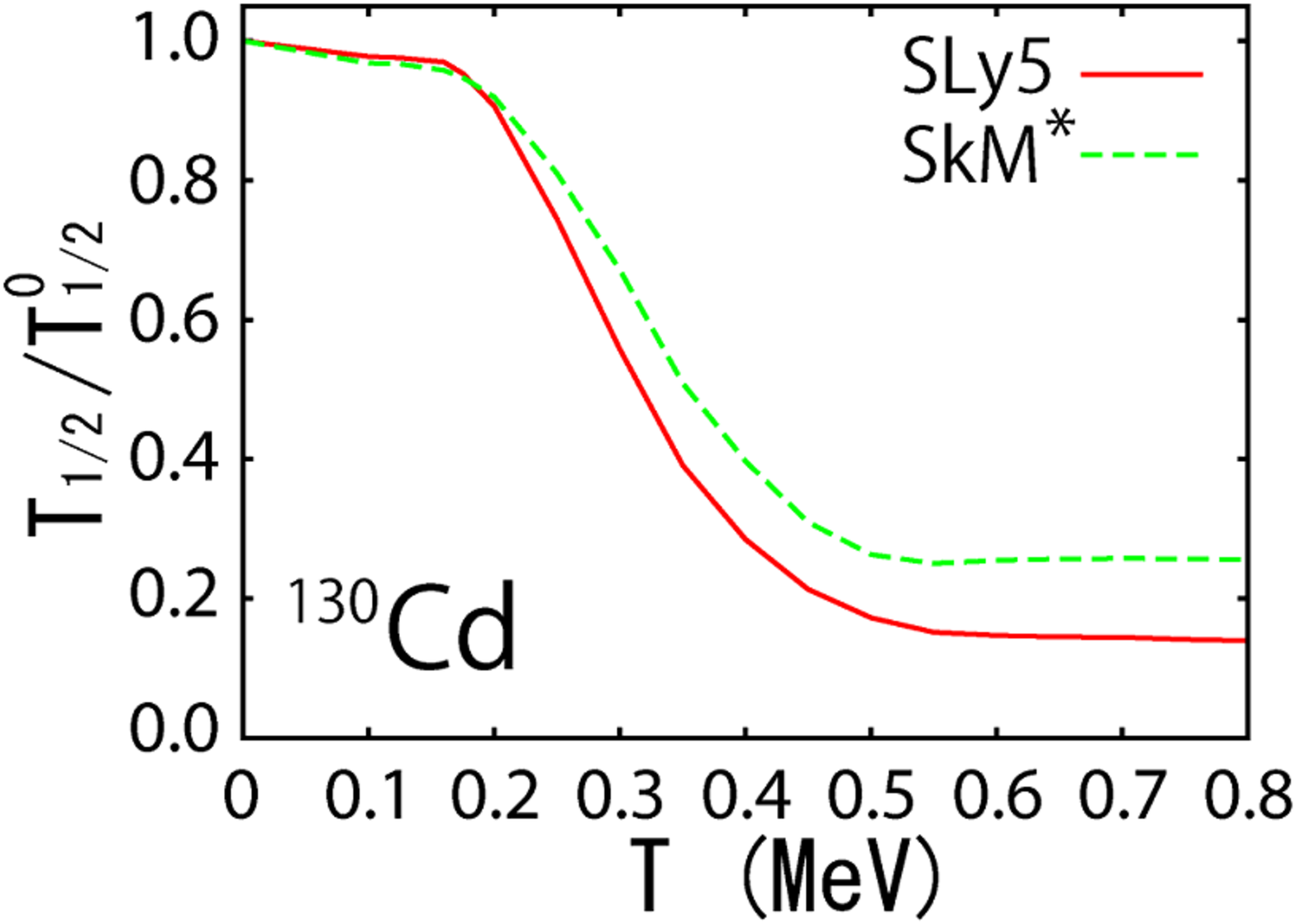}
\end{tabular}
\caption{(Color online)
The $\beta$-decay half-life $T_{1/2}$ normalized to 
that at zero-temperature, $T_{1/2}^0$.
The solid and the dashed lines show the results with 
the SLy5 and SkM$^*$ parameter sets, respectively.}
\label{Half-lives}
\end{center}
\end{figure*}

Figure \ref{Half-lives} shows the $\beta$-decay half-lives normalized to 
that at zero temperature, $T_{1/2}/T_{1/2}^0$, as a function 
of temperature $T$. 
In order to check the parameter set dependence of the Skyrme functional,
the figure also shows the results with 
the SkM$^*$ parameter set \cite{SkM}.
One sees that, as the temperature increases, 
the $\beta$-decay half-life first 
decreases gradually for all the nuclei we study for both the parameter sets.

One can also see that the temperature dependence 
is the stronger for the larger atomic number. 
For instance, at $T=0.8$ MeV, 
the ratio $T_{1/2}/T_{1/2}^0$ is around 0.2 
for $^{130}$Cd both for the SkM$^*$ and SLy5, 
while it is about 0.9 for $^{120}$Sr.
This behavior is related with the number of the GT peaks.
Figure \ref{GTstrengthbeta} indicates that 
the number of GT peaks decreases gradually with the atomic number.
This is due to the difference between the proton and neutron Fermi surfaces.
For $^{130}$Cd, 
the number of GT peaks is only two at $T$=0,
and the thermal effects are relatively large.
On the other hand, 
the effects are less significant for $^{122}$Zr 
because there are already many strengths at $T=0.0$ MeV.

For $^{122}$Zr, $^{124}$Mo, and $^{126}$Ru, the half-lives begin to increase 
at temperature around $T=0.6-0.7$ MeV.
This is related to the temperature dependence of $E_{\rm{QRPA}}$, which 
also influences the phase space factor in Eq. \eqref{PhaseSpace}.
That is, 
when $E_{\rm{QRPA}}$ is large,
the phase space factor is also large, resulting in 
a large $\beta$-decay rate ({\it i.e.,} a short half-life). 
Since the excitation energy from the ground state, 
$E_i^{\rm{ex}} \equiv E_i-E_0$ and $E_{\rm{QRPA}}$ now depend on the temperature, 
the $\beta$-decay half-life may not behave in a simple way as a function of $T$.

A simple estimate of the thermal effect on 
the $\beta$-decay half-life can be made by disregarding the temperature 
dependence of $S^-(E_m)$ in 
Eq. \eqref{betadecayrate} except for the the thermal population probability 
of excited levels. 
We approximate the population probability
by the Boltzmann statistical factor, $\exp(-E_i^{\rm{ex}}/T)$. 
The $\beta$-decay rate is then proportional to 
\begin{equation}
e^{-(E_i-E_0)/T}\times E_e \sqrt{E_e^2-m_e^2}(E_i-E_m^*-E_e)^2,
\end{equation}
which has a local maximum at $E_i=2T+E_m^*+E_e$.
That is, 
the $\beta$-decay rate is large when the condition 
$E_i^{\rm{ex}} = E_i-E_0\sim 2T+E_e-(E_0-E_m^*)$ is satisfied. 
On the other hand, 
it is hindered considerably in the case of 
$E_i^{\rm{ex}} \ll 2T+E_e-(E_0-E_m^*)$.

\begin{figure}[t]
\begin{center}
\includegraphics[width=.44\textwidth,clip]{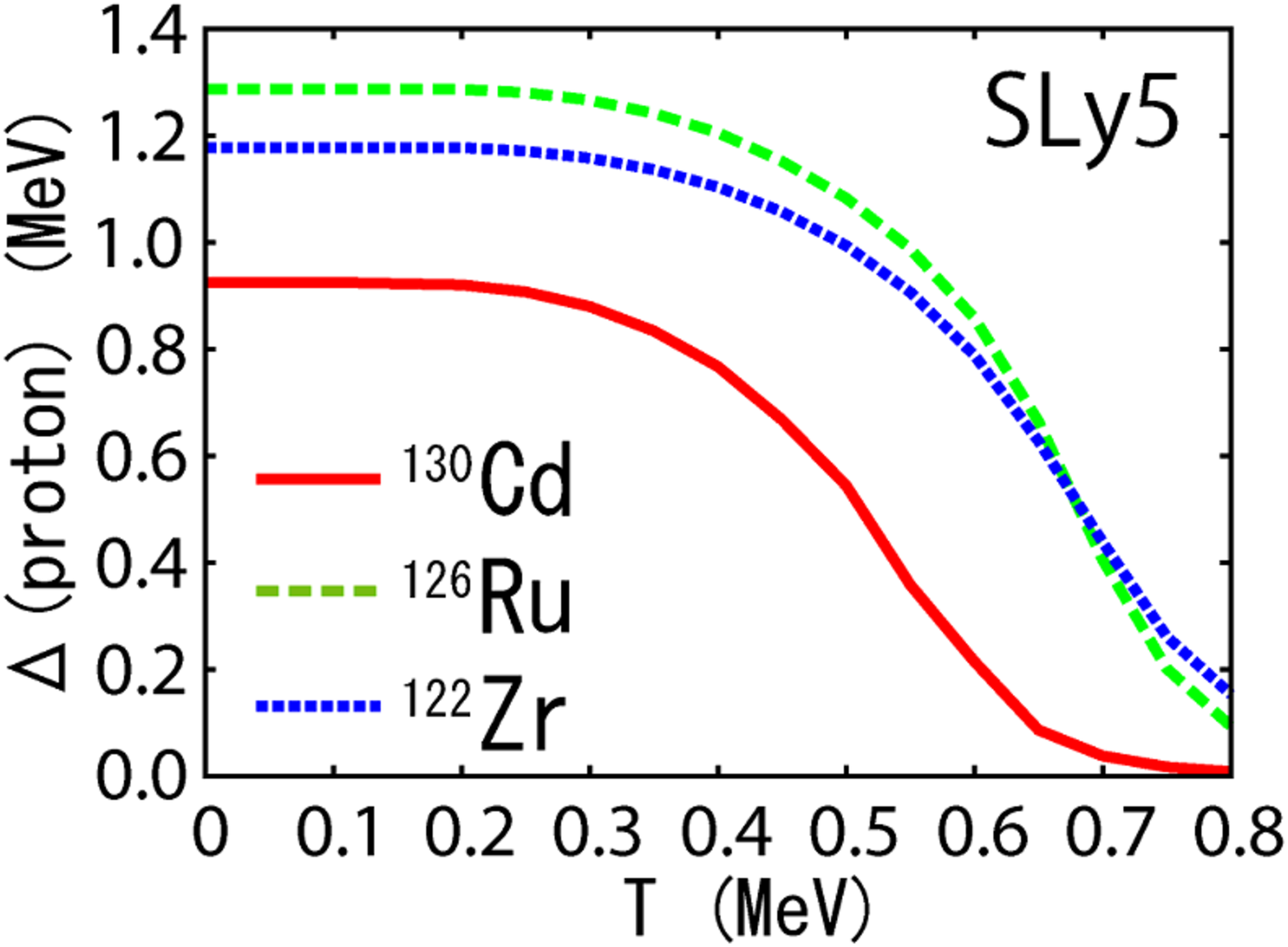}
\includegraphics[width=.44\textwidth,clip]{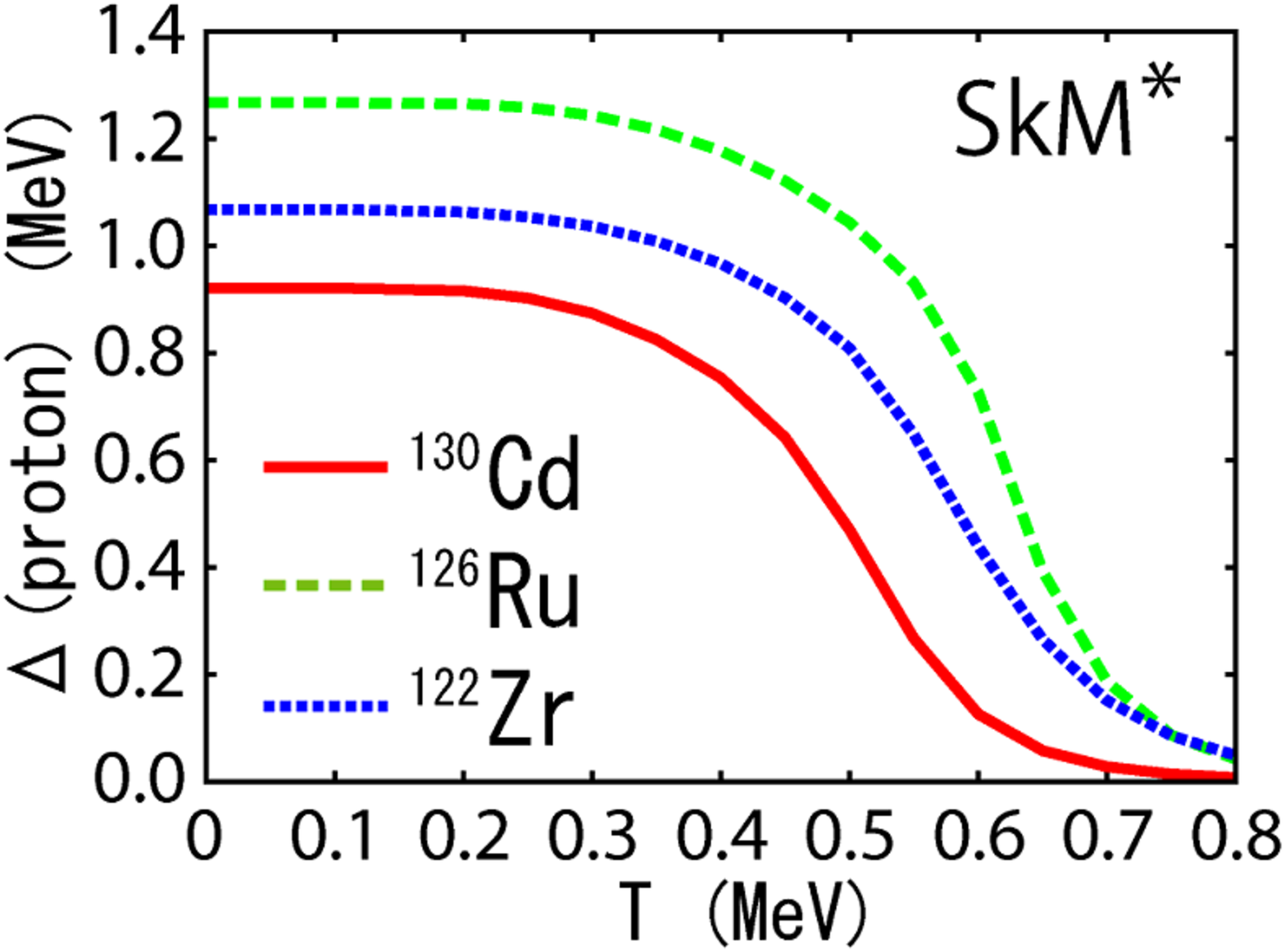}
\caption{(Color online) 
The average proton pairing gap as a function of temperature 
for the $^{130}$Cd, $^{126}$Ru, 
and $^{122}$Zr nuclei. The top and bottom panels are results 
for the SLy5 and SkM$^*$ parameter sets, respectively.} 
\label{dlt}
\end{center}
\end{figure}
\begin{figure}[t]
\begin{center}
\includegraphics[width=.44\textwidth,clip]{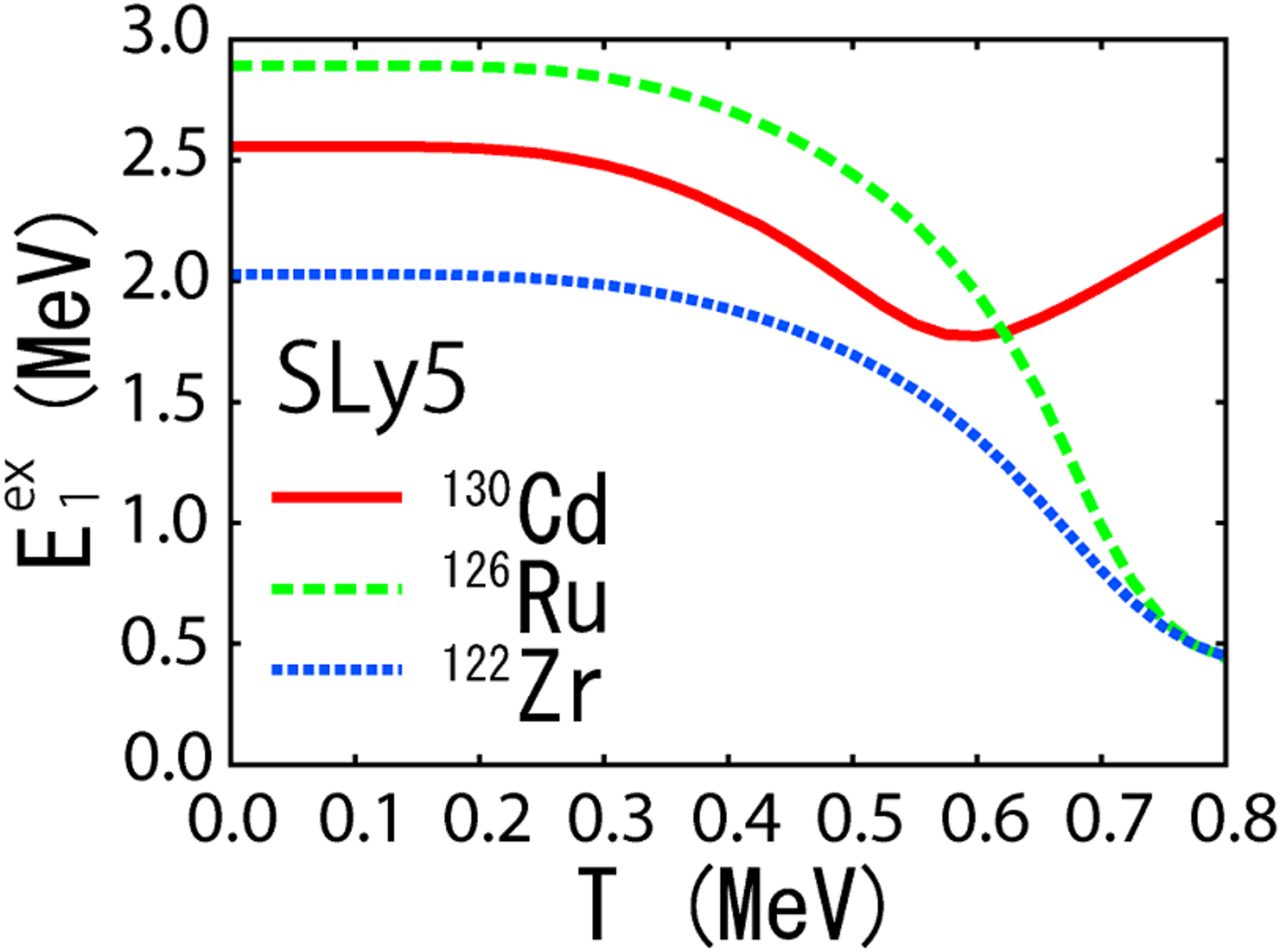}
\includegraphics[width=.44\textwidth,clip]{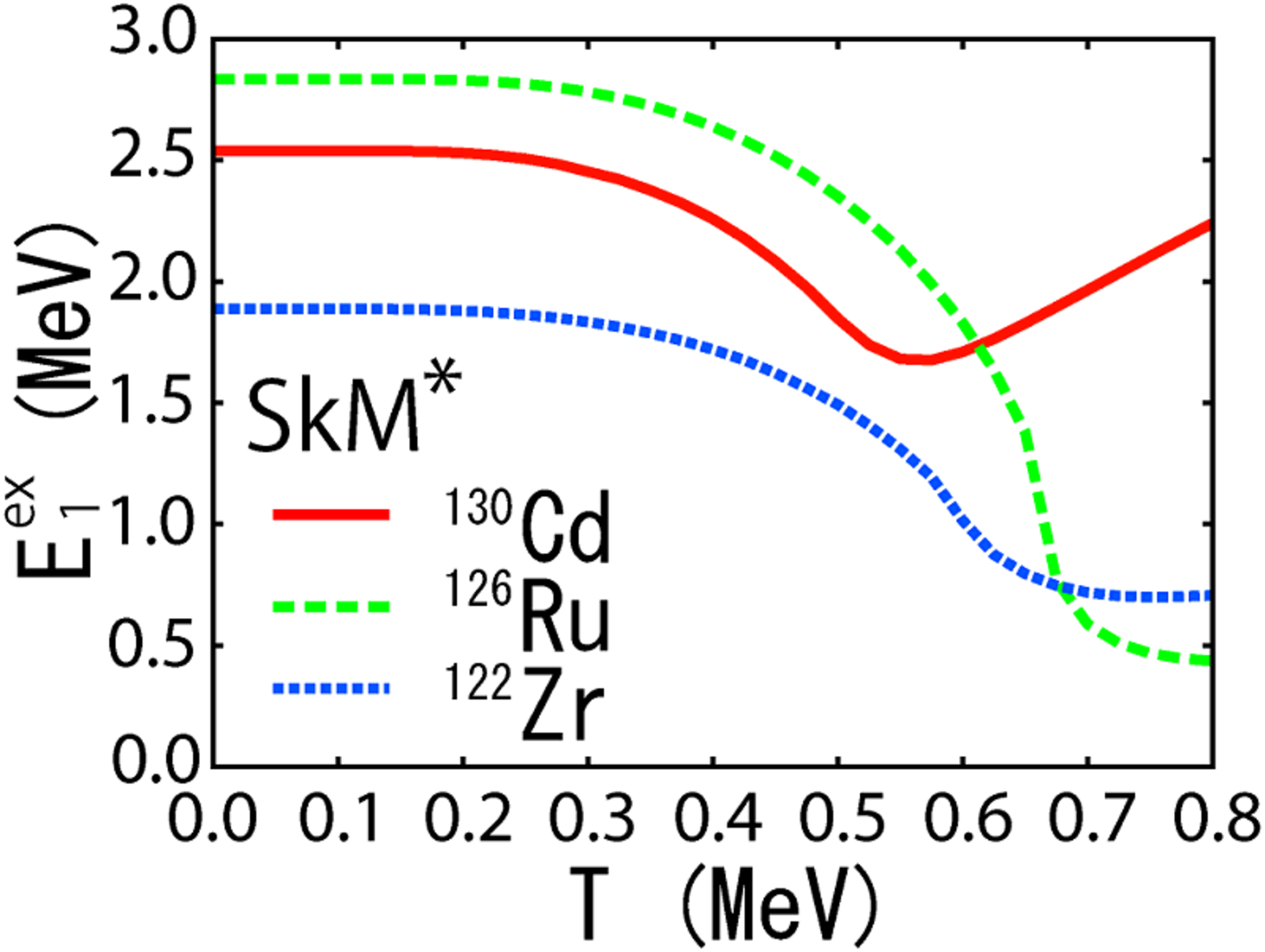}
\caption{(Color online) 
Same as Fig. \ref{dlt}, but for the unperturbed energy 
of the first excited state, estimated in the BCS approximation.}
\label{bcs2qp}
\end{center}
\end{figure}

In order to investigate the temperature dependence of 
$E_i^{\rm{ex}}$, 
we plot the average pairing gaps $\langle \Delta \rangle$ 
in Figure \ref{dlt} and
the unperturbed energy of the first excited state $E_{1}^{\rm{ex}}=E_1-E_0$, 
evaluated with the two-quasi-particle energy in the BCS approximation,  
in Figure \ref{bcs2qp}. 
The top and bottom panels show the results of 
the SLy5 and SkM$^*$ parameter sets, respectively. 
One sees that the pairing gaps begin to decrease
significantly at temperatures of about $T=0.50-0.65$ MeV
({\it i.e.,} the pairing phase transition).
Likewise, the energy of the first excited state, $E_i^{\rm{ex}}$, also 
decreases rapidly at similar temperatures. 
For $^{126}$Ru and $^{122}$Zr, it eventually becomes 
less than $2T$ at high temperatures. 
This should be intimately related to the increase of 
the $\beta^-$-decay half-lives for these nuclei at high temperatures.
On the other hand,
$E_{1}^{\rm{ex}}$ for $^{130}$Cd 
is much less sensitive to the temperature, 
as this nucleus is in the neighborhood of the double magic nucleus $^{132}$Sn. 
As a consequence, its $\beta$-decay half-life 
monotonically decreases as a function of temperature. 
Note that the critical temperature for the pairing phase transition is 
lower for SkM$^*$ as compared to SLy5.
This fact leads to the result that the $\beta$-decay half-lives start 
increasing earlier for SkM$^*$ compared to SLy5, 
as can be seen in Figure \ref{Half-lives}.

\section{CONCLUSION}

We have assessed the thermal effects on $\beta$-decay half-lives
with astrophysical interests 
for even-even isotones with the neutron magic number $N=82$. 
For this purpose,  
we have adopted the finite temperature QRPA method on top of 
the finite temperature Skyrme-Hartree-Fock+BCS method.  
We have used the $t_0$ and $t_3$ terms of the Skyrme force 
for the particle-hole residual interaction, 
and a $\delta$-type interaction for the proton-neutron 
particle-particle channel in the QRPA formalism.

We have calculated the Gamow-Teller strengths in the temperature range 
from $T=0.0$ to $0.8$ MeV.
At finite temperatures, 
new peaks appear in the strength function 
due to the transitions from the excited states.
From the calculated Gamow-Teller strengths, 
we have evaluated the $\beta$-decay half-lives. 
As the temperature increases, 
the $\beta$-decay half-life decreases gradually for all the nuclei which 
we have studied.
We have also found that the temperature dependence appears 
more strongly for nuclei with a larger atomic number.
We have argued that 
this is related to the number of GT peaks in the strength 
function, determined mainly by the
difference between the proton and neutron Fermi surfaces.
We have also found that the $\beta$-decay half-life 
begins to increase at $T>0.6-0.7$ MeV for open-shell nuclei as a 
consequence of a peculiar temperature dependence of 
the energy of the first excited state due to the pairing phase transition.

From our results, 
we conclude that the thermal effect on 
the $\beta$-decay half-life is negligible 
at the standard r-process temperature, 
which is considered to be approximately less than $0.2$ MeV,
at least for even-even $N=82$ isotones. 
It would be an interesting future problem 
to extend the present calculations to odd-mass nuclei, in which 
the energy of the first excited state 
is in general much smaller than that in even-even nuclei and thus 
a larger thermal effects may be expected.

\section*{Acknowledgment}

We thank G. Col\`o, T. Kajino, and H. Sagawa for useful discussions.
This work was supported by the GCOE programme 
``Weaving Science Web beyond Particle-Matter Hierarchy'' at Tohoku University, 
and by the Japanese
Ministry of Education, Culture, Sports, Science and Technology
by Grant-in-Aid for Scientific Research under
the program number 19740115.

\end{document}